\documentclass{iau}

\usepackage{amsmath}
\usepackage{graphicx}
\usepackage{multirow}

\begin{document}

\lefttitle{Vinicius Branco~{\it et. al}}
\righttitle{Spectrophotometry in the integrated light of multiple populations in globular clusters}

\jnlPage{1}{7}
\jnlDoiYr{2025}
\doival{10.1017/xxxxx}
\volno{395}
\pubYr{2025}
\journaltitle{Stellar populations in the Milky Way and beyond}

\aopheadtitle{Proceedings of the IAU Symposium}
\editors{J. Mel\'endez,  C. Chiappini, R. Schiavon \& M. Trevisan, eds.}

\title{Spectrophotometry in the integrated light of multiple populations in globular clusters}

\author{V. Branco\textsuperscript{$\star$}, A. Lançon, P. Coelho, G. Costa, T. Dumont, L. Martins, P. Prugniel, F. Martins, C. Charbonnel, and A. Palacios.}
\affiliation{\textsuperscript{$\star$}Universidade de São Paulo, IAG, Brazil.\\\textsuperscript{$\star$}Université de Strasbourg, CNRS, Observatoire astronomique de Strasbourg, France.\\vbranco@usp.br}


\begin{abstract}
There is vast evidence from observations of multiple stellar populations (MPs) in globular clusters (GCs). 
To explore the issue theoretically, this work considers two subsolar metallicities, two ages, and two initial abundance patterns: a first population of standard $\alpha$-enhanced metal mixture stars and a second stellar population displaying C-N and Na-O anticorrelations chemical abundance patterns, along with an enhanced helium fraction.
Analysing the predictions for these extreme compositions, we provide insights into the observability of not-resolved MPs into individual stars of GCs. 
We use colours and spectrophotometric indices measurable with modern facilities (e.g. {\em Euclid}, LSST, DES, JWST).
\end{abstract}

\begin{keywords}
atlases; 
globular clusters: general; 
stars: atmospheres; stars: Population II
\end{keywords}

\maketitle

\section{Introduction}

Most Galactic globular clusters (GCs) harbour multiple populations of stars (MPs) that are composed of at least two generations: the first generation is characterised by a standard $\alpha$-enhanced metal mixture (hereafter ``\textbf{1P}''), as observed in field halo stars of the Milky Way, and the second generation display C-N and Na-O anticorrelations chemical abundance patterns in combination with an enhanced helium fraction (hereafter ``\textbf{2P}'').
Adequate collections of stellar spectra are needed to describe the effect of these changes in the stellar abundance on the integrated light of GCs.
Following up on our previous work \citep{Branco2024}, in which we presented a grid of synthetic stellar spectra built upon observed stars of Galactic GCs, we present our preliminary analysis of stellar populations built upon isochrones.

\section{Synthetic stellar spectra with abundances representative of GC stars}

Based on {\tt STAREVOL} tracks described in \cite{Chantereau2015}, \cite{Costa2024} computed a new collection of isochrones that 
extend to the end of the thermally pulsing AGB.
These isochrones consider Fast Rotating Massive Stars (FRMS) as the main polluters of the 2P. 
Isochrones based on a super-massive star scenario for GC formation and 2P pollution are under construction.

\subsection{Isochrones for cluster sub-populations}

We computed stellar atmosphere models and synthetic spectra with \textsc{ATLAS12} and \textsc{SYNTHE} \citep{Kurucz2005} for eight isochrones with two chemical surface abundance assumptions, representatives of the first and second generations. Hence, two chemical abundance patterns were considered:

\begin{itemize}

    \item \textbf{1P}: a {standard} metal mixture with solar-scaled compositions and initial He mass fraction;    
    \item \textbf{2P}: a second generation whose metal composition has depleted C, O, Mg, Li, Be, and B, and enriched He, N, Na, and Al.  
\end{itemize}

Table \ref{tab:gc_ssp} summarizes the grid of isochrones considered in this work.

\begin{table}[!h]
\centering
\
\caption{{Fundamental parameters of the eight isochrones of our dataset.}}

\begin{tabular}{|c|c|c|c|c|} 

\toprule      
 \textbf{[Fe/H]} &  \textbf{Z} &  \textbf{Y}  &  \textbf{Ages} (Gyr) & \textbf{ref. NGC} \\

\midrule 

$-1.52$, $-1.43$ & $0.0009$ & $0.248$, $0.400$ &  $9.5$, $13.0$ & $6752$\\
$-1.14$, $-1.05$ & $0.0020$ & $0.249$, $0.400$ &  $9.5$, $13.0$ & $2808$\\
       
\bottomrule
\end{tabular} 
\label{tab:gc_ssp}
\end{table} 

All surface abundances used to compute the atmospheric models were obtained from the isochrones, from $^\mathrm{1}$H to $^\mathrm{37}$Cl, including the proportions of some iron-peak elements. The wavelengths range from 0.24 to 3 $\mu$m, an interval over which numerous spectrophotometric indices are sensitive to the surface abundance variations associated with MPs. 

\section{Integrated spectra}

\begin{figure}[!ht]
\centering
  \includegraphics[width=\columnwidth]{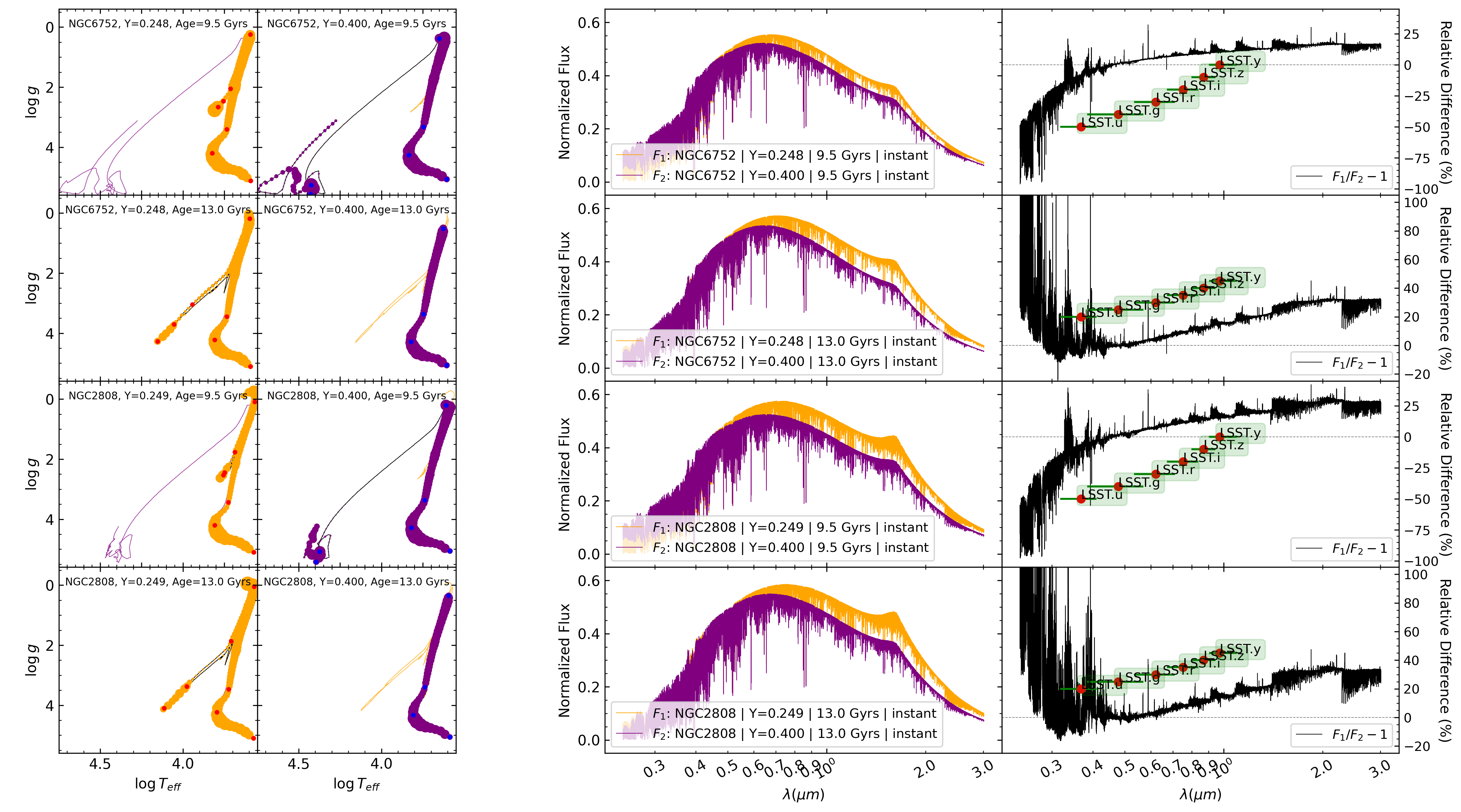}
  \caption{\emph{Left}: Isochrones of the eight SSPs. Point sizes represent the light contribution to the SED. \emph{Right:} SEDs from the isochrones of the 1P (orange) and 2P (purple) of each SSP (rescaled $\lambda F_{\lambda}$) and the relative differences between 1P ($F_1$) and 2P ($F_2$), with overplotted LSST filters.}
  \label{fig:isochrones-spectra}
\end{figure}

Adopting a standard IMF \citep{Kroupa2001}, we populated the isochrones and computed the integrated spectra of Single Stellar Populations (SSPs).
Along with the isochrones, Figure \ref{fig:isochrones-spectra} illustrates the 1P and 2P spectra for the two metallicities and ages, alongside the relative differences between each sub-population. 
The presence or absence of red post-RGB phases determines the spectra at optical and near-IR wavelengths; in the near-UV, blue horizontal branches and hot post-RGB phases (AGB manqué) affect the spectra.

With these new calculations, we can assess how 2P abundances impact SSP spectra via changes in the stellar tracks (evolutionary effect) and surface abundances (spectral effect).
Evolutionary effects of $\alpha$-enhancement and MPs in the integrated light have been discussed previously in \cite{Coelho2007} and \cite{Coelho2011}. Our ingredients in this work update those previous discussions with newer synthetic spectra and the FRMS scenario.

\section{Integrated spectrophotometry}

\begin{figure}[!ht]
    \centering

    \includegraphics[width=0.38\columnwidth]{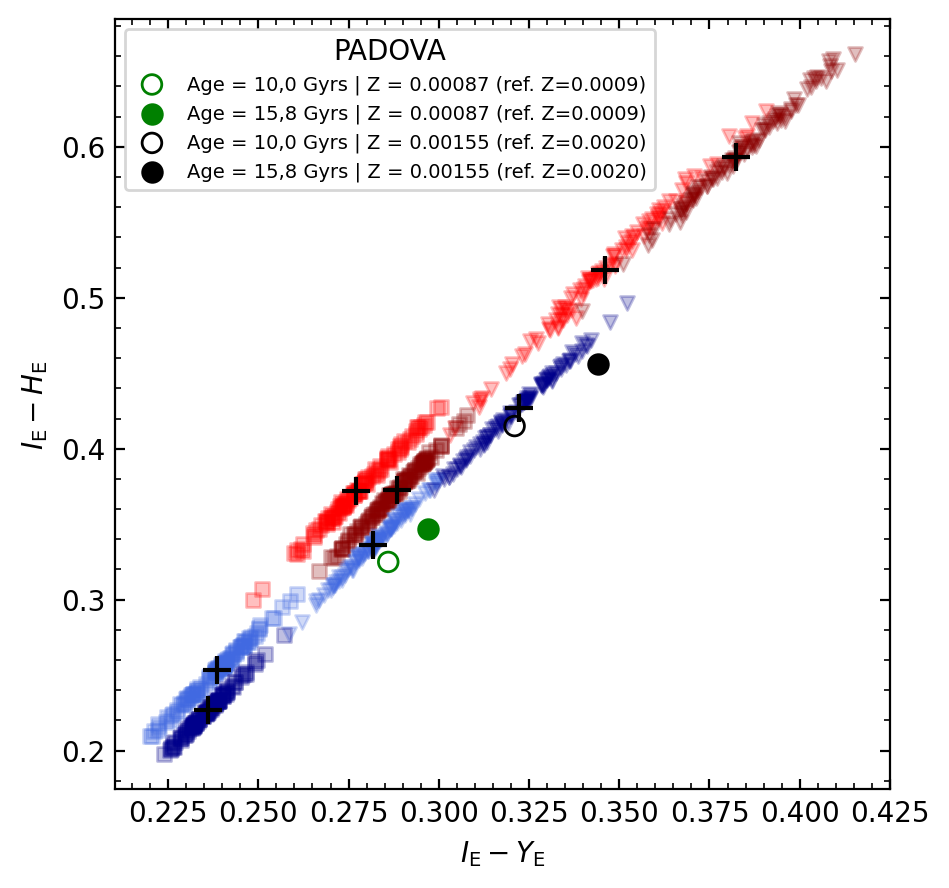}
    \includegraphics[width=0.38\columnwidth]{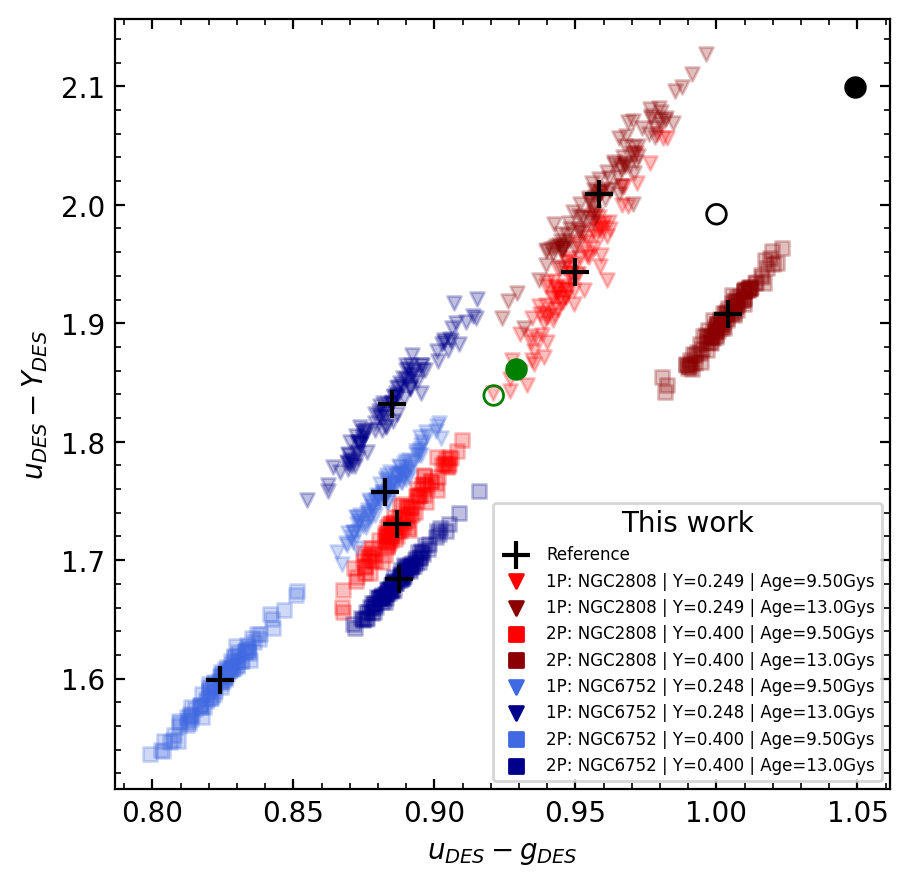}
    \\
    \includegraphics[width=0.38\columnwidth]{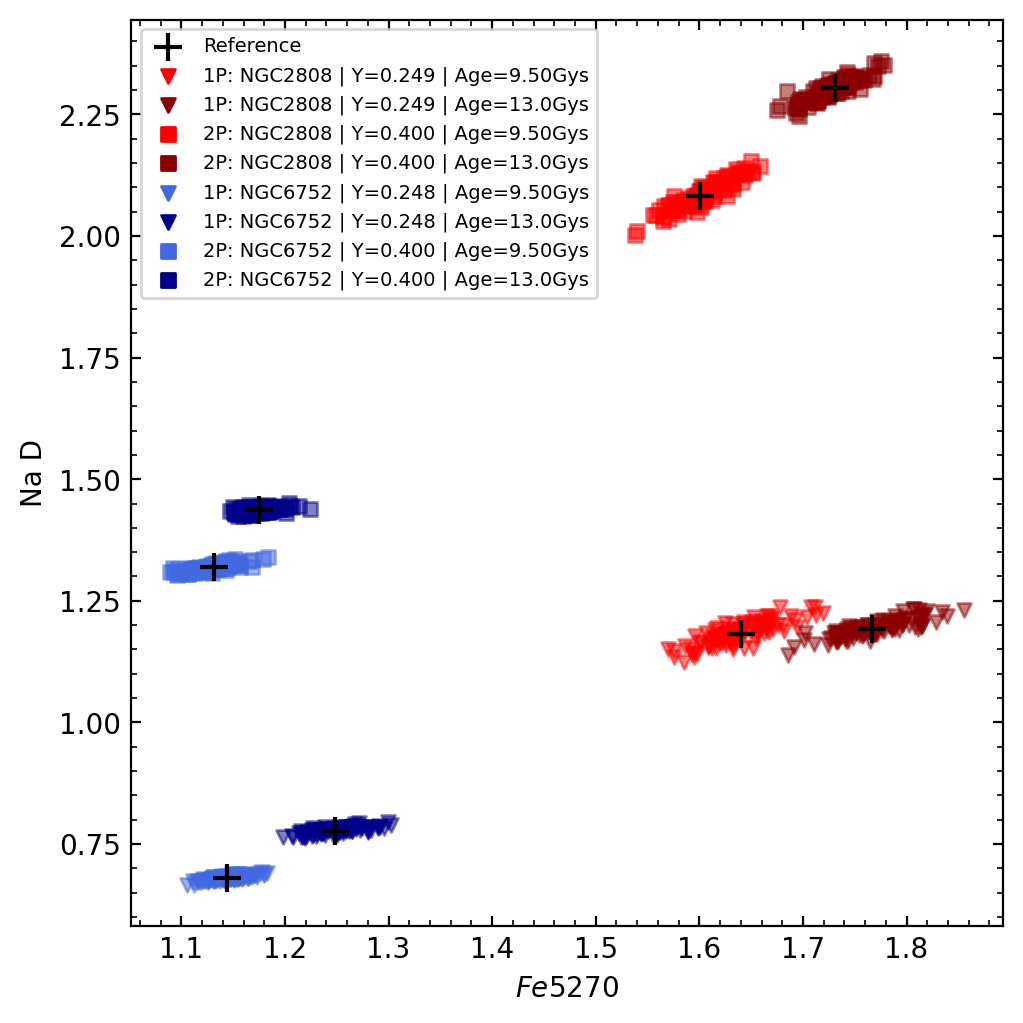}
    \includegraphics[width=0.38\columnwidth]{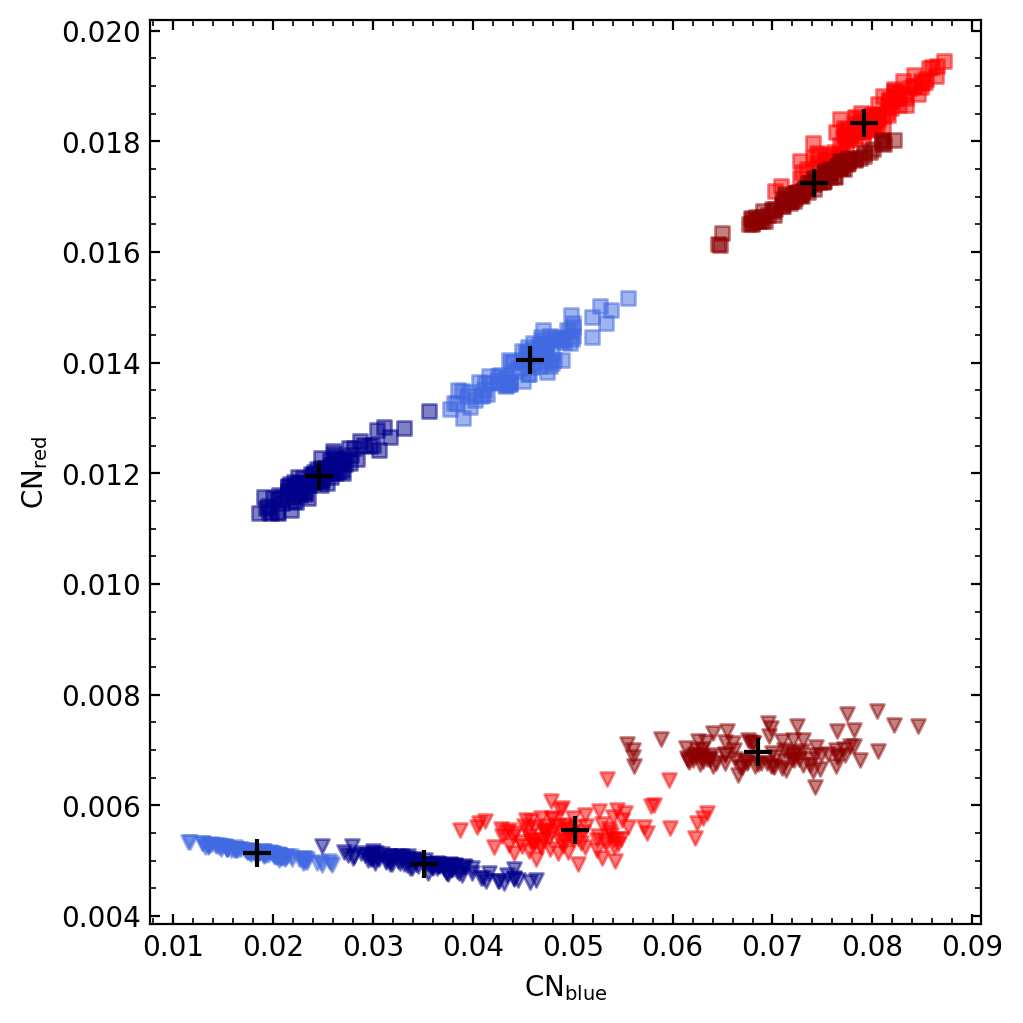}

  \caption{Spectrophotometric indices of the stochastic populations. \emph{Top:} Colour-colour diagrams of the eight SSPs, based on {\em Euclid} (left) and Dark Energy Survey filters (right). Colours for a few Padova isochrones (1P) are added. \emph{Bottom:} Spectral indices of the eight SSPs.}
  \label{fig:spectrophotometry}
\end{figure}

The effects of age, metallicity, and the 1P/2P composition on selected 
spectrophotometric indices are shown in Figure \ref{fig:spectrophotometry} measured in bootstrap populations, i.e. one hundred stochastic populations each containing one million stars drawn from the IMF. 
The photometric indices show good general agreement with predictions from the default isochrones of the Padova group (1P only). Still, it can be seen that the new dimension (1P \textit{vs} 2P) adds extra complexity and extra degeneracies.
The spectral indices sensitive to the MP effects are mostly consistent with the results we obtained in \citet{Branco2024}. 
The enhancement in Na in population 2P is reflected in the higher Na\,D spectral index. 
Enhanced N in 2P-stars leads to generally stronger CN bands.
However, compared with 1P, age affects the 2P differently, inverting the trend. This may be due to the old 2P isochrones in the HRDs stopping at the RGB tip, thus affecting the blue region of the spectra with less light contribution.

\section{Conclusions \& Perspectives}

Integrated spectra are sensitive to the shape of the isochrones in HR diagrams and the different surface abundances of 1P and 2P stars. This manifests in the low-resolution SEDs, molecular bands involving C, N, and O, and line spectra.
The (artificial) lack of hot post-RGB or post-AGB phases in some isochrone calculations impacts the UV region.
Hence, while $\mathrm{CN}_\mathrm{red}$ is consistent with our previous work, $\mathrm{CN}_\mathrm{blue}$ needs to be re-examined. 
The next generation of isochrones will include extra mixing processes such as thermohaline mixing, leading to larger differences.

\small{
    \emph{Acknowledgements:} 
    \footnotesize{This study was financed in part by CAPES (88887.580690/2020-00), and by ANR, France, under project POPSYCLE (ANR-19-CE31-0022). We also acknowledge support from CNPq (200928/2022-8 and 310555/2021-3) and FAPESP (2021/08813-7).}
    }

\bibliographystyle{iau} 
\bibliography{iau} 

\begin{thebibliography}{}

\bibitem[{Branco} et~al., 2024]{Branco2024}
{Branco}, V., {Coelho}, P. R.~T., {Lan{\c{c}}on}, A., {Martins}, L.~P., \& {Prugniel}, P. 2024, Synthetic stellar spectra for studying multiple populations in globular clusters. extended grid, and the effects on the integrated light.
\newblock \aap, 687, A142.

\bibitem[{Chantereau} et~al., 2015]{Chantereau2015}
{Chantereau}, W., {Charbonnel}, C., \& {Decressin}, T. 2015, {Evolution of long-lived globular cluster stars. I. Grid of stellar models with helium enhancement at [Fe/H] = -1.75}.
\newblock \aap, 578, A117.

\bibitem[{Coelho} et~al., 2007]{Coelho2007}
{Coelho}, P., {Bruzual}, G., {Charlot}, S., {Weiss}, A., {Barbuy}, B., \& {Ferguson}, J.~W. 2007, {Spectral models for solar-scaled and {\ensuremath{\alpha}}-enhanced stellar populations}.
\newblock \mnras, 382, 2, 498-514.

\bibitem[{Coelho} et~al., 2011]{Coelho2011}
{Coelho}, P., {Percival}, S.~M., \& {Salaris}, M. 2011, {Chemical Abundance Anticorrelations in Globular Cluster Stars: The Effect on Cluster Integrated Spectra}.
\newblock \apj, 734, 1, 72.

\bibitem[{Costa} et~al., 2024]{Costa2024}
{Costa}, G., {Dumont}, T., {Lan{\c{c}}on}, A., {Palacios}, A., {Charbonnel}, C., {Prugniel}, P., {Ekstrom}, S., {Georgy}, C., {Branco}, V., {Coelho}, P., {Martins}, L., {Borisov}, S., {Voggel}, K., \& {Chantereau}, W. 2024, {He-enriched STAREVOL models for globular cluster multiple populations: Self-consistent isochrones from ZAMS to the TP-AGB phase}.
\newblock \aap, 690, A22.

\bibitem[{Kroupa}, 2001]{Kroupa2001}
{Kroupa}, P. 2001, {On the variation of the initial mass function}.
\newblock \mnras, 322, 2, 231-246.

\bibitem[{Kurucz}, 2005]{Kurucz2005}
{Kurucz}, R.~L. 2005, {ATLAS12, SYNTHE, ATLAS9, WIDTH9, et cetera}.
\newblock Memorie della Societa Astronomica Italiana Supplementi, 8, 14.

\end{thebibliography}










\end{document}